\documentclass{IEEEtran4PSCC}

\ifCLASSINFOpdf
  \usepackage[pdftex]{graphicx}
\else
  \usepackage[dvips]{graphicx}
 \fi

\usepackage{comment}
\usepackage[cmex10]{amsmath}

\usepackage{mathtools}

\usepackage{tikz}
\usepackage[shortlabels]{enumitem}
\usepackage{pgf,tikz,pgfplots}
\pgfplotsset{compat=1.13}
\usepackage{mathrsfs}
\usetikzlibrary{arrows}
\usepackage{multicol}
\usepackage{subfigure}
\usepackage{bbm}
\usepackage{amssymb,amsmath,amsthm}
\usepackage[ruled,vlined,linesnumbered]{algorithm2e}
\usepackage{hyperref}

\newtheorem{theorem}{Theorem}

\newcommand{\pw}{\mathbbm{P}_{\omega}}

\allowdisplaybreaks

\makeatletter
\let\old@ps@headings\ps@headings
\let\old@ps@IEEEtitlepagestyle\ps@IEEEtitlepagestyle
\def\psccfooter#1{%
  \def\ps@headings{%
    \old@ps@headings%
    \def\@oddfoot{\strut\hfill#1\hfill\strut}%
    \def\@evenfoot{\strut\hfill#1\hfill\strut}%
  }%
  \def\ps@IEEEtitlepagestyle{%
    \old@ps@IEEEtitlepagestyle%
    \def\@oddfoot{\strut\hfill#1\hfill\strut}%
    \def\@evenfoot{\strut\hfill#1\hfill\strut}%
  }%
  \ps@headings%
}
\makeatother

\psccfooter{%
    \parbox{\textwidth}{\hrulefill \\ \small{21st Power Systems Computation Conference} \hfill \begin{minipage}{0.2\textwidth}\centering \vspace*{4pt} \includegraphics[scale=0.06]{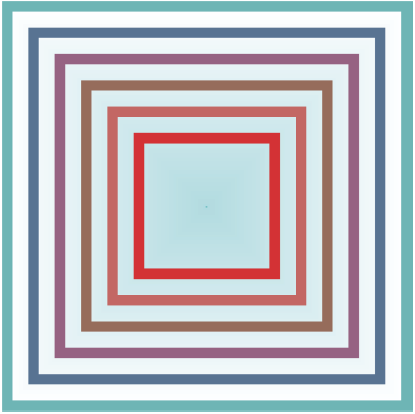}\\\small{PSCC 2020} \end{minipage} \hfill \small{Porto, Portugal --- June 29 -- July 3, 2020}}%
}

\begin{document}
%
\title{Stochastic AC Optimal Power Flow:\\ A Data-Driven Approach}

\author{
\IEEEauthorblockN{Ilyes Mezghani}
\IEEEauthorblockA{CORE, UCLouvain \\
Louvain-la-Neuve, Belgium\\
ilyes.mezghani@uclouvain.be}
\and
\IEEEauthorblockN{Sidhant Misra and Deepjyoti Deka}
\IEEEauthorblockA{Theory Division, Los Alamos Nationial Laboratory\\
Los Alamos, NM, USA\\
\{sidhant, deepjyoti\}@lanl.gov}
}


\newcommand{\algo}{\textsc{DDS-OPF}}

\maketitle

\begin{abstract}
There is an emerging need for efficient solutions to stochastic AC Optimal Power Flow ({AC-}OPF) to ensure optimal and reliable grid operations in the presence of increasing demand and generation uncertainty. This paper presents a highly scalable data-driven algorithm for stochastic AC-OPF that has extremely low sample requirement. The novelty behind the algorithm's performance involves an iterative scenario design approach that merges information regarding constraint violations in the system with data-driven sparse regression. Compared to conventional methods with random scenario sampling, our approach is able to provide feasible operating points for realistic systems with much lower sample requirements. Furthermore, multiple sub-tasks in our approach can be easily paralleled and based on historical data to enhance its performance and application. We demonstrate the computational improvements of our approach through simulations on different test cases in the IEEE PES PGLib-OPF benchmark library.
\end{abstract}



\section{Introduction}
Modern power systems are faced with significant uncertainty in power generation and demand. This is due to increasing integration of renewable energy resources like wind and solar, and growth of demand side participation and distributed energy resources at the sub-transmission and distribution levels. As a result, uncertainty management has become a critical component in the operational planning stage, where generators and controllable elements must be dispatched in a way that the system remains within its safety limits despite uncertain fluctuations. In the literature, the issue is addressed by considering variants of the optimal power flow (OPF) problem that incorporate the effect of uncertainty. These formulations take the form of either a stochastic or robust optimization problem, where a limit on some chosen measure of risk is explicitly enforced. 

All uncertainty-aware OPF formulations pose significant computational challenges, most of which can be traced back to the non-linear implicit nature of the AC power flow equations. The two primary challenges are 
(i) quantifying the effect of uncertainty on the system -- it is difficult to precisely express the variation of the dependent physical quantities in the system such as voltage magnitudes and line currents as a function of the uncertainty, and (ii) formulating a sufficiently compact optimization that integrates the uncertainty quantification while still being tractable. These challenges have been echoed in several recent publications on the topic \cite{roald2017chance}, \cite{Muhlpfordt2019TPWRS}, and several solution approaches have been proposed. Broadly, we can classify these approaches into two types (i) approximations to the AC power flow equations, and (ii) Monte Carlo methods.

\textbf{Power flow approximations:} These class of approaches aim at simplifying the task of uncertainty quantification by full or partial approximations to the power flow equations. These include linear approximations of the power flow such as the DC approximation \cite{stott2009dc} and a first order Taylor expansion \cite{tcns}. Using these approximations greatly improves tractability, in particular for risk metrics that can be expressed as a convex program. Many publications \cite{roald2017chance,lorca2017robust,molzahn18,venzke2018,Muhlpfordt2019TPWRS,dall2017chance} have attempted to incorporate the AC power flow equations. In \cite{roald2017chance}, only a partial linearization is considered, where all nominal quantities follow the full non-linear AC-PF while the effect of uncertainty is expressed via linearization. The resulting method is much more accurate than full linearization, but can lose fidelity when the magnitude of the uncertainty is large.  More recently, an approach based on polynomial chaos expansion \cite{Muhlpfordt2019TPWRS} has been proposed that is highly accurate but computationally challenging. In summary, approaches based on power flow approximation trade-off accuracy for scalability -- the brief review mentioned above cites methods with \emph{high scalability - low accuracy} to \emph{low scalability - high accuracy}. 

\textbf{Monte Carlo methods:} These methods quantify the effect of uncertainty on the system by solving the power flow equations for a large number of realizations drawn from the uncertainty distribution. When the number of samples used is sufficiently large, Monte Carlo provides excellent accuracy. The primary challenge however lies in integrating Monte Carlo into an optimization formulation without exploding the size of the problem {and the associated computational time}. In this context, the most widely used approach is the so-called \emph{scenario approach} where an extended OPF problem is formulated by incorporating a specified number of scenarios from the distribution, and robustness to each scenario is enforced via constraints. Several theoretical results \cite{calafiore2006scenario,vrakopoulou2013probabilistic} (primarily for convex formulations with chance constraints) provide guidelines on how many random samples should be used to achieve the desired probability of constraint violation. The main drawback of the approach is that random sampling based approaches, specifically for non-linear and non-convex optimal power flow problems, quickly result in the optimization problem becoming computationally intractable for practical cases.

\textbf{Contribution:} In this paper, we adopt the scenario-based approach described above. However {instead of including scenarios collected randomly}, we use system knowledge and data-driven tools to drastically reduce the number of scenarios required to solve the problem. This keeps the primary benefits of the scenario approach -- accurate uncertainty quantification, agnosticism to uncertainty distribution, etc., while significantly boosting its scalability. Our approach is an advanced iterative procedure similar to scenario generation algorithms common in problems such as power systems expansion planning \cite{mashayekh2017security}. The algorithm iteratively adds more scenarios to the scenario-based OPF until a security criterion, assessed by a sufficient number of Monte Carlo samples, is satisfied. Note that since the assessment of scenarios \emph{does not} involve solving the OPF, scalability is not compromised and the procedure can heavily exploit availability of parallel computing capabilities. Following the scenario assessment, what is added back to the OPF, in each iteration, is a well-chosen subset of `modified' scenarios. To determine the `modified' scenarios, we first develop metrics for sub-selecting a very small portion of critical scenarios based on constraint violation. Second, we use regularized linear regression to identify the directions of uncertainty that are the most adversarial for each violated constraint. We then boost the chosen critical scenarios along the directions identified and add them back to the OPF formulation for the next iteration. We show through several case studies that this data-driven program significantly reduces the scenario size requirements over vanilla random sampling - with $\sim 30$ scenarios we are able to find a secure solution to the stochastic OPF for the large $1354$ bus system. In summary, our contribution in this paper is a suite of data-driven tools to efficiently solve the stochastic OPF problem with a scenario-based approach. The features used in our scenario selection procedure can be tuned based on historical knowledge/expertise available with an operator. The code-base used for our implementation is being released with this paper for testing/validation by the research community. 


\section*{Notation}
\underline{Sets}
\begin{itemize}[leftmargin=*]
    \item $\mathcal{B}$, set of buses.~~ $\mathcal{L}$, set of lines.~~$\mathcal{G}$, set of generators.
    \item $\mathcal{PV}$ / $\mathcal{PQ}$, set of $PV$ / $PQ$ buses. $\mathcal{G}_{sl}$, slack bus.
    \item $\Omega$, uncertainty set. $\Omega_N$, finite set of $N$ scenarios $\omega \in\Omega$.
\end{itemize}
\underline{Parameters}
\begin{itemize}[leftmargin=*]
    \item $P_i$ / $Q_i$, real / reactive demand at bus $i$.
    \item $G_i$ / $B_i$, shunt conductance/ susceptance at bus $i$. 
    \item $G_{ij}$ / $B_{ij}$, conductance / susceptance on line $(i,j)$.
    \item $\underline{x}$ / $\overline{x}$, lower / upper capacity limit on variable $x$.
    \item $S_{ij}$, flow limit on line $(i,j)$.
    \item $\mu_i^p(\omega)$ \ $\mu_i^q(\omega)$, real / reactive demand fluctuation at bus $i$ under scenario $\omega$.
    \item $c_g(.)$, cost of generator $g\in \mathcal{G}$, assumed to be convex quadratic.
\end{itemize}
\underline{Variables}
\begin{itemize}[leftmargin=*]
    \item $p_i$ / $q_i$, real / reactive injection at node $i$.
    \item $f^p_{ij}$ / $f^q_{ij}$, real / reactive power flow on line $(i,j)$.
    \item $v_i$ / $\theta_i$, voltage magnitude / angle at node $i$.
    \item $p^0_g$ / $v^0_g$, generation / voltage set point at $PV$ bus $g$. 
\end{itemize}


\section{Problem Formulation}  \label{sec:formulation}
In this section, we provide details of modeling a power system subject to uncertain power injections, the corresponding generation recourse policy and details of the stochastic OPF formulation. 

\subsection{Power system under uncertainty} \label{subsec:uncertainty_modeling}
We consider a power network and denote the set of buses by $\mathcal{B}$ and the transmission lines by $\mathcal{L}$. Without loss of generality, in the presentation that follows we assume at most one generator and one load per bus, and that the net power injection at each bus is subject to uncertainty. Consider an uncertainty realization $\omega$ in possibly unknown/non-parametric set $\Omega$. The set of power flow equations under uncertainty $\omega$ are given by
\begin{subequations}  \label{eq:PFE_uncertainty}
\begin{align}
&\sum_{(i,j) \in \mathcal{L}} f^p_{ij}(\omega) = p_i(\omega) - (P_i+\mu^p_i(\omega)) - G^s_iv_i^2(\omega) \quad \forall i \in \mathcal{B}\label{eq:RealPowerBalance}\\
&\sum_{(i,j) \in \mathcal{L}} f^q_{ij}(\omega) = q_i(\omega) - (Q_i+\mu_i^q(\omega)) + B^s_iv_i^2(\omega) \quad \forall i \in \mathcal{B}\label{eq:ReactivePowerBalance}\\
& f^p_{ij}(\omega) = G_iv_i^2(\omega) - G_{ij}v_i(\omega)v_j(\omega) \cos(\theta_i(\omega)-\theta_j(\omega))\notag\\&\qquad - B_{ij}v_i(\omega)v_j(\omega) \sin(\theta_i(\omega)-\theta_j(\omega)) \qquad \forall (i,j) \in \mathcal{L}\label{eq:FpDef}\\
& f^q_{ij}(\omega) = -B_iv_i^2(\omega) + B_{ij}v_i(\omega)v_j(\omega) \cos(\theta_i(\omega)-\theta_j(\omega))\notag\\&\qquad - G_{ij}v_i(\omega)v_j(\omega) \sin(\theta_i(\omega)-\theta_j(\omega)) \qquad \forall (i,j) \in \mathcal{L}\label{eq:FqDef}
\end{align}
\end{subequations}
In \eqref{eq:PFE_uncertainty}, $\mu_i^p(\omega), \mu_i^q(\omega)$ denotes the active and reactive power fluctuations at bus $i$, under uncertainty $\omega$. All other variables in the system are explicitly expressed as a function of the uncertainty realization. \\

\noindent \textit{Recourse Model:} For a non-zero realization of uncertainty, the generators in the system must adjust their generation to maintain total power balance and feasibility. We use an affine policy representing the automatic generation control (AGC) that is representative of current power system operation \cite{roald2013analytical}.
\begin{subequations}  \label{eq:recourse}
\begin{align}
  & p_g(\omega) = p_g^0 + \left(\sum_{i \in \mathcal{B}}\mu_i^{p}(\omega)\right)\alpha_g, \ \forall g \in \mathcal{PV}, \ \forall \omega \in \Omega \label{eq:defPg}\\
  & v_g(\omega)=v_g^0, \ \forall i \in \mathcal{PV}, \ \forall \omega \in \Omega \label{eq:defV}
\end{align}
\end{subequations}
Equation~\eqref{eq:defPg} shows the linear adjustment in the active power generation of generator $g$ from its nominal value of $p_g^0$ as a fraction of the total power mismatch $\sum_{i \in \mathcal{B}}\mu_i^{p}(\omega)$ caused by the uncertainty, according to its participation factor $\alpha_g$. In this paper, we consider the participation factors to be given and fixed. For simplicity, we assume $\alpha_g=\frac{1}{|\mathcal{G}|}$, although this specific choice is not relevant for our method. Equation~\eqref{eq:defV} says that the voltage magnitudes at $PV$ buses are kept constant during operation, and is in accordance with current practice.


\subsection{Stochastic Optimal Power Flow Formulations}
In this section, we present the stochastic optimal power flow problem in a generic form. Since our solution approach involves a Monte Carlo in-the-loop validation step, we have the flexibility to handle a variety of such formulations. We state the set of inequality constraints in the OPF representing the standard safety limits {on line flows, phase angle difference at neighboring buses, and bus injections and voltages} that need to be enforced.
\begin{subequations}  \label{eq:safety_limits}
\begin{align}  
\Gamma_{\text{OPF}} = & \left\{ \ (p,q,f^p,f^q,v,\theta) \mid \right. \nonumber \\
& \quad (f^p_{ij})^2 + (f^q_{ij})^2 \leq S_{ij}^2 \qquad \forall (i,j) \in \mathcal{L} \label{eq:LineLimFrom}\\ 
& \quad \underline{\theta_{ij}} \leq \theta_i - \theta_j\leq \overline{\theta_{ij}} \qquad \forall (i,j) \in \mathcal{L} \label{eq:boxTheta}\\
& \quad \underline{p} \leq p \leq \overline{p}, \ \underline{q} \leq q \leq \overline{q}, \ \underline{v} \leq v \leq \left. \overline{v} \ \right\} \label{eq:box}
\end{align}
\end{subequations}
In the above definition, $\Gamma_{\text{OPF}}$ denotes the set of all power flow solutions that satisfy the safety limits given in \eqref{eq:safety_limits}. \\

\noindent \textit{Dependent and independent variables:} For clarity of exposition, we first specify which variables in the stochastic OPF are controllable/independent and which variables are dependent. Suppose that the nominal values of generation {$p^0$} and voltages {$v^0$} at the PV buses have been determined. Assume that for each realization of the uncertainty $\omega$, the generators react according to the recourse policy in \eqref{eq:recourse}. Then given $\omega$, Equations~\eqref{eq:recourse}, fully determine the active power generation and voltage magnitude $p_i(\omega), v_i(\omega)$ at all $PV$ buses. The (known) functions $\mu_i^p(\omega), \mu_i^q(\omega)$ fully determine all real and reactive power injections $p(\omega), q_i(\omega)$ at the $PQ$ buses. Once these variables are specified, we are in the standard Power Flow setting, and the set of equations in \eqref{eq:PFE_uncertainty} fully specify the value of the rest of the variables -- $q_i(\omega),\theta_i(\omega)$ at the $PV$ buses, and $v_i(\omega),\theta_i(\omega)$ at the $PQ$ buses, and all line flows $f_{ij}^p(\omega),f_{ij}^p(\omega)$. We summarize this functional dependence using the following notation:
\begin{align}  \label{eq:functional_dependence}
  (p(\omega),q(\omega),f(\omega),v(\omega),\theta(\omega)) = \text{PF}\left(p^0, v^0,\omega; \alpha\right).
\end{align}
A stochastic optimal power flow problem in generic form corresponds to finding a set of nominal set point values for the active power generation $p_g^0$ and voltage magnitude $v^0$ such that the total generation cost is minimized, and some stochastic measure of power flow violation for a given uncertainty distribution is below a required limit $\epsilon$. This is made precise in the formulation below:
\begin{subequations}  \label{eq:stochastic_opf}
\begin{align}
\min_{p^0, v^0} \quad & \sum_{g \in \mathcal{G}} c_g(p^0_g) \label{eq:obj}\\
\mbox{s.t.} \quad & \mathcal{SV} = \mathbf{E}_{\pw}\left[\mathcal{V}\bigg( \text{PF}\left(p_g^0, v^0,\omega; \alpha\right) ,\Gamma_{\text{OPF}}  \bigg)\right] \leq \epsilon. \label{eq:stochastic_violation}
\end{align}
\end{subequations}
Equation~\eqref{eq:obj} specifies the objective that minimizes the total \emph{nominal} generation cost. This is for simplicity. It is possible to incorporate the cost of reserves in a straightforward way. Equation~\eqref{eq:stochastic_violation} enforces that some \emph{stochastic violation measure} is bounded. The stochastic violation measure $\mathcal{SV}$ is the expectation of some violation measure $\mathcal{V}()$ with respect to $\pw$ which denotes the probability distribution of the uncertainty $\omega$. The violation measure $\mathcal{V}()$ is a function of the uncertainty dependent power flow variables (first argument) and the feasibility/safety region (second argument), and is used to quantify how far the uncertain power flow variables are from the feasible region.  Note that the generic formulation in \eqref{eq:stochastic_opf} includes common cases, such as, 

\textit{Chance Constrained OPF (CCOPF)\cite{bienstock2014chance}:} This formulation enforces that the probability of constraint violation is smaller than a specified value and corresponds to 
\begin{align}  \label{eq:speical_cc_opf}
    &\mathcal{V}\bigg( \text{PF}\left(p_g^0, v^0,\omega; \alpha\right) ,\Gamma_{\text{OPF}}  \bigg)  \nonumber \\
    &\qquad \qquad = \mathbbm{1}\big(p(\omega),q(\omega),f(\omega),v(\omega),\theta(\omega) \notin \Gamma_{\text{OPF}}\big),  \\
   &\mathcal{SV} = \pw\bigg( p(\omega),q(\omega),f(\omega),v(\omega),\theta(\omega) \notin \Gamma_{\text{OPF}} \bigg),
\end{align}
where $\mathbbm{1}()$ denotes the indicator function. 
By a linear combination of the different constraints in \eqref{eq:safety_limits},  Equation~\eqref{eq:speical_cc_opf} can be converted to individual, and weighted Chance Constrained OPF \cite{roald2015optimal}

Closed-form analytic expressions for the stochastic constraint in Equation~\eqref{eq:stochastic_violation} are not easy to derive for the AC-PF model under general uncertainty distributions. To overcome intractability, data driven scenario OPF can be formulated. 

\subsection{Scenario OPF (S-OPF)}\label{subsec:scenario_approach_recap}
Scenario approach \cite{calafiore2006scenario,vrakopoulou2013probabilistic} collects a set $\Omega_N$ of $N$ random samples for the uncertainty $\omega \in \Omega$. By definition, the base case $\omega = 0$ is included in set $\Omega_N$, and the user is assumed to have access to a scenario generation/sampling process (from historical data or otherwise). We then solve an OPF problem with hard feasibility constraints for each selected scenario as denoted below. 
\begin{subequations}  \label{eq:basic_scenario_opf}
\begin{align}
\min_{p^0, v^0} \quad & \sum_{g \in \mathcal{G}} c_g(p^0_g) \label{eq:obj1}\\
\mbox{s.t.} \quad & \forall \omega_i \in \Omega_N, \  \text{PF}\left(p_g^0, v^0,\omega_i; \alpha\right) \in \Gamma_{\text{OPF}}\label{eq:sample_violation}
\end{align}
\end{subequations}
By ensuring feasibility for a large-enough and representative sample set $\Omega_N$, S-OPF can indirectly guarantee the stochastic violation constraint Equation~\eqref{eq:stochastic_violation} 
Theoretical bounds on the size of the sample set necessary to ensure $\mathcal{SV}() \leq \epsilon$ and {related design of box constraints exist for convex optimization problems \cite{calafiore2006scenario,margellos2014road}}, but are not generalizable to AC-OPF. As demonstrated later, the number of samples to ensure low stochastic violation quickly grows. This makes the standard S-OPF in \eqref{eq:basic_scenario_opf} computationally intractable for realistically sized test cases. {Existing scenario selection methods pick a sub-set of scenarios from the ones available, randomly \cite{sequential} or by minimizing a inter-distribution distance such as Wasserstein metric \cite{growe2003scenario}. Similarly, mixed-integer programs have been proposed to pick a sub-set of scenarios inside chance-constrained optimization \cite{sabbir_reduction}. However the number of selected scenarios necessary, or the mixed-integer programs themselves, still involve a large computational requirement for AC- OPF. In this paper, we take a different approach where system knowledge and data-driven techniques are combined to design (not just select) strategic scenarios that lead to a drastically more efficient scenario OPF. }

\section{Data-driven Scenario OPF} \label{sec:algorithm}
The overarching goal of our approach is to determine an optimized scenario set $\Omega_N$ of far lesser cardinality, compared to random sampling, so that a tractable scenario OPF solution with stochastic violations below prescribed threshold can be be determined. We propose an algorithm called \algo{}:
\begin{algorithm}
\SetAlgoLined
\textbf{Initialization:}  Solve S-OPF \eqref{eq:basic_scenario_opf} using rated loads $P_i, Q_i~ (|\Omega_N|=1)$ to get $(p^0, v^0)$ \;
\textbf{Monte Carlo:}  Sample a set of permissible scenarios $\mathcal{S}$ of size $S$ according to $\pw$. Solve the PF with recourse for each scenario \;
\textbf{Stopping criterion check:} Check if the \emph{estimated} stochastic violation measure $\tilde{\mathcal{SV}}$ is below pre-selected threshold, $\tilde{\mathcal{SV}} = \frac{1}{S}\sum_{i=1}^S \mathcal{V}(*,\omega_i) < \tau$. If yes, {\bf exit} \;
\textbf{Scenario construction:}  Use data-driven methods to design $K < S$ scenarios to add to $\Omega_N$ \;
\textbf{Update:} Compute new solution $(p^0, v^0)$ for \eqref{eq:basic_scenario_opf} with $\Omega_N$. Go to Step {\bf Monte Carlo} \;
{\bf return} $p^0, v^0$.
\caption{\algo{}}
\label{alg:overall_algorithm}
\end{algorithm}

The threshold $\tau$ used in \algo{} is selected based on the properties of the stochastic violation measure $\mathcal{SV}()$, pre-fixed $\epsilon$ bound (see \eqref{eq:stochastic_violation}), and the confidence requirement. Theoretical confidence bound on the solution for selected $\tau$ is given in Section~\ref{subsec:confidence_bounds}. All numerical experiments considered in this paper focus on the case when $\mathcal{SV}()$ corresponds to the probability of constraint violation. For that, the estimated stochastic violation measure $\tilde{\mathcal{SV}}$ simply corresponds to the fraction of samples in $\mathcal{S}$, for which the constraints are violated. 

The rest of the section is focused on describing the critical Step~$\mathbf{4}$ in \algo{}. To guide intuition, we use computations on the \texttt{pglib\_opf\_case73\_ieee\_rts} test-case in the OPF Power Grid Library \cite{pglib2019}. This case has $73$ buses, $120$ lines and $51$ loads. We assume that $\pw$ is a uniform distribution within a box, i.e., for each load $i \in \mathcal{B}$ we have $\mu_i^p(\omega) \sim \mathcal{U}\left[-0.03P_i,0.03P_i \right]$ and $\mu_i^q(\omega) \sim \mathcal{U}\left[ -0.03Q_i,0.03Q_i\right]$, where $\mathcal{U}$ denotes the uniform distribution, and $P_i,Q_i$ are the rated active and reactive demands.

\begin{figure}
\centering
	\includegraphics[width=.5\textwidth]{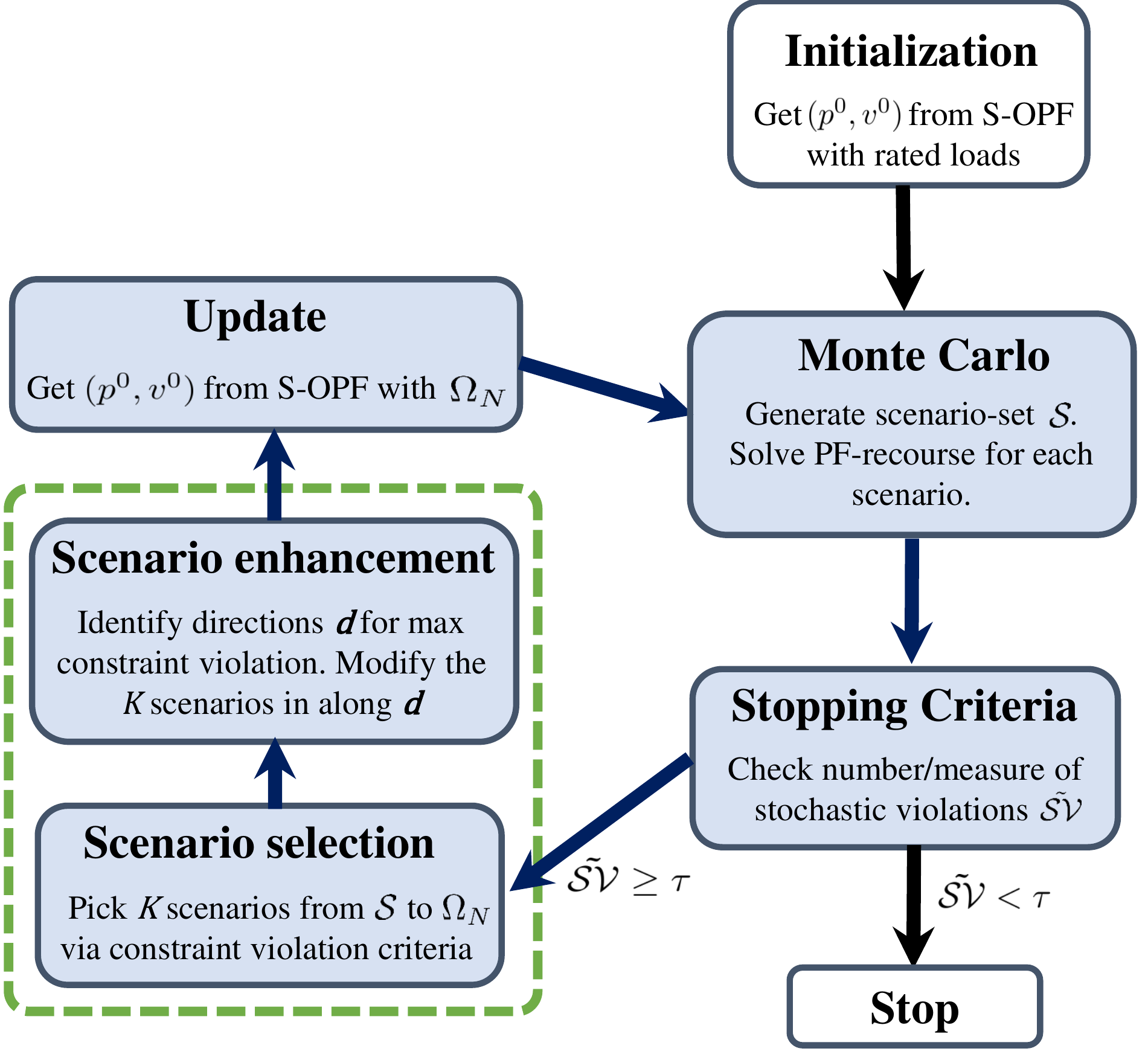}
\caption{Schematic of \textbf{DDS-OPF}. The scenario construction sub-steps are highlighted within the green box.}
\label{fig:general_idea}
\vspace{-4pt}
\end{figure}

Table~\ref{tab:TE_results_first} shows the performance of the vanilla scenario approach where randomly drawn samples are included in the set $\Omega_N$.
\begin{table}[ht]
  \centering
  \caption{Feasibility on $1,000$ out-of-sample scenarios for DDS-OPF with randomly sampled $\Omega_N$ with $K=10$, for $73$-bus test system.}
  \begin{tabular}{|l|c|c|c|c|c|c|}
  \hline
  $|\Omega_N|-1$ & 1 & 10 & 20 & 30 & 50 \\
  \hline
  $P^{1000}_{vio}$ & 100\% & 59.5\% & 25.0\% & 32.3\% & 12.2\%\\
  \hline
  {Cost} &  {1.904e5} &  {1.948e5} &  {1.948e5} &  {1.948e5} &  {1.948e5}  \\
  \hline
  \end{tabular}
  \label{tab:TE_results_first}
\end{table}
Note that non-trivial number of violations are still obtained despite $50$ scenarios. This high sample requirement prevents tractability for realistic test-cases. To improve over random sampling, our proposed \emph{scenario construction} in Step~$\mathbf{4}$ includes 2 key sub-steps: 
\begin{enumerate}[(a)]
  \item \emph{PF-aware scenario selection}: We use prioritization metrics to down-select \emph{dominant} scenarios.
  \item \emph{Data-driven scenario enhancement}: For scenarios selected in (a), we identify critical directions that maximize their effect on S-OPF, and modify them (stretch or squeeze) along these directions before adding to $\Omega_N$.
\end{enumerate}
A schematic representation of our overall approach is shown in Fig.~\ref{fig:general_idea}. In what follows, we describe in detail, the motivation and important features of scenario construction sub-steps and use the $73$-bus test system to demonstrate improvements.

\subsection{PF-aware scenario selection} \label{sec:pf-aware_sel}
A random scenario, that is already feasible for the current solution $(p^0, v^0)$, is less likely to be effective for feasibility improvement than a scenario that has multiple constraint violations during recourse. We use information about constraint violations to sub-select a small number of \emph{dominant} scenarios from set $\mathcal{S}$ in Step~$\mathbf{2}$ to add to the scenario set $\Omega_N$. Fortunately, the infeasible scenarios and their corresponding constraint violations are already acquired while validating the performance of $(p^0, v^0)$ in Step~$\mathbf{3}$. 

\subsubsection{Dominant scenario selection} We consider three different prioritization criteria: 
\begin{itemize}
    \item \emph{Maximum violation (MV).} Scenarios having the largest constraint violation, measured relative to bound value. 
    \item \emph{Number of constraints (NC).} Scenarios violating the maximum number of constraints.
    \item \emph{Hybrid.} Scenarios that have the highest ${weight}_s= \frac{MV_s}{\max\limits_{s'\in \mathcal{S}} MV_{s'}} + \frac{NC_s}{\max\limits_{s'\in \mathcal{S}} NC_{s'}}$, where $MV_s$ is the largest violation of a constraint, and $NC_s$ is the number of constraints violated, by scenario $s$.
\end{itemize}
We avoid selecting a new scenario that violates the same set of constraints as a previously selected (dominating) scenario. Such avoidance ensures that a greater proportion of constraint violations are represented in $\Omega_N$. 
\subsubsection{Batch size selection} While the prioritization criteria rank the scenarios according to their dominance, the number of samples $K$ that are added back to $\Omega_N$ still needs to be decided and can have a significant impact on overall efficiency. When $K$ is too small the total number of iterations can be large since we are adding very little information to the problem in each iteration. On the other hand, when $K$ is too large, the size of the resulting S-OPF can quickly make it intractable. Through multiple numerical experiments, we confirm that a batch size of $5$ provides the right trade-off across a variety of test cases. 

\subsubsection{Results for $73$-bus case} 
\begin{figure}
  \centering
  \includegraphics[scale=0.17]{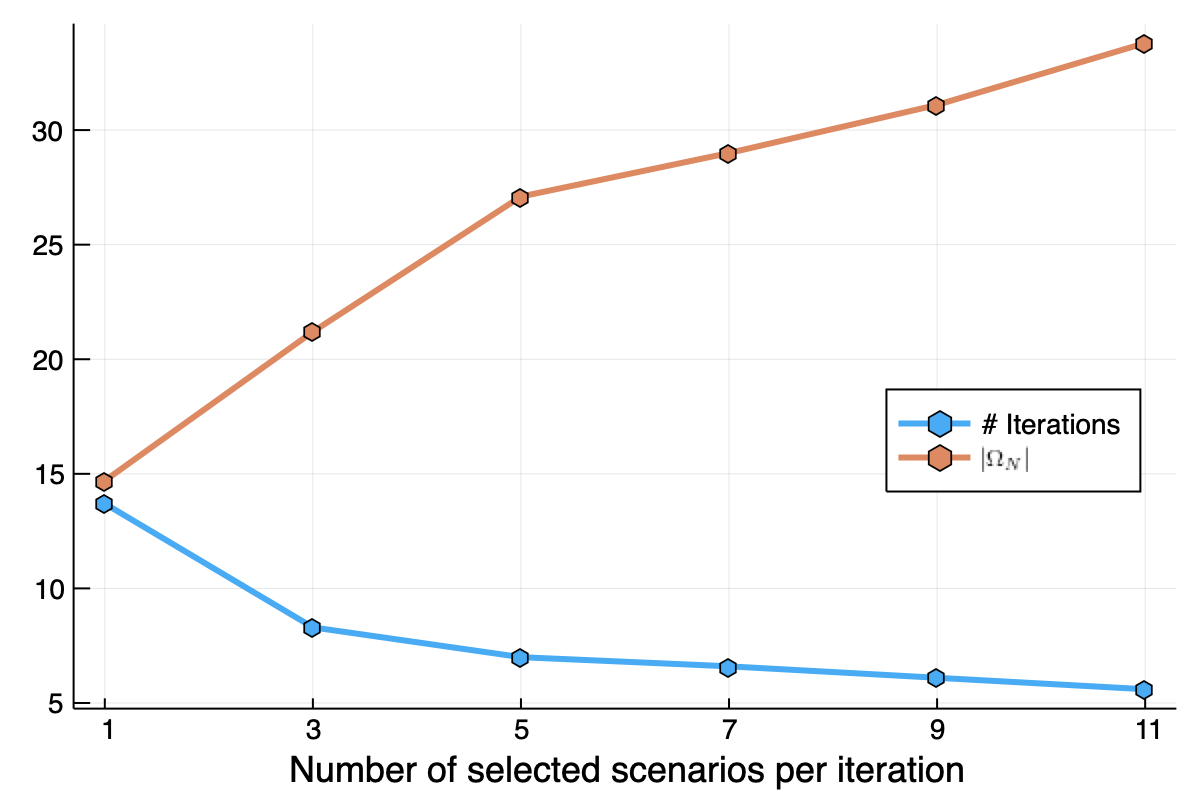}
  \caption{Number of iterations and final size of $\Omega_N$ when the algorithm has converged for different choices for $K$, number of selected scenarios per iteration, for the $73$ bus system. These are average numbers out of 10 runs of DDS-OPF.}
  \label{fig:choice_n}
\end{figure}

We use dominant scenario selection in \algo{} with $S = 1000$, $K=5$ and $\tau = 0$ and show the results in Table \ref{tab:TE_data_driven_selection}. 
\begin{table}[!htb]
  \centering
  \caption{Feasibility on $1,000$ out-of-sample scenarios for DDS-OPF with scenario selection with $K=5$, for $73$-bus test system.}
  \begin{tabular}{|l|c|c|c|c|c|c|}
  \hline
  Policy & \# Iterations & $|\Omega_N|$ & $P^{1000}_{vio}$\\
  \hline
  MV & 5 & 20 & 0.1 \%\\
  \hline
  NC & 7 & 28 & 0 \%\\
  \hline
   Hybrid & 8 & 29 & 0 \%\\
  \hline
  \end{tabular}
  \label{tab:TE_data_driven_selection}
\end{table}
Compared to the results for the random sampling in Table~\ref{tab:TE_results_first}, we have significantly improved performance for each of the three proposed criteria. Indeed, at most $29$ scenarios are able to reduce the number of infeasible scenarios to almost zero on out-of-sample testing.
Figure~\ref{fig:choice_n} shows how the number of iterations and total scenario size $\Omega_N$ changes for various choices of $K$, justifying our choice of $K=5$. Note that the number of final scenarios is often less than $\#iterations * K + 1$, since in each iteration, only one of multiple scenarios that violate the same set of constraints, is added to $\Omega_N$. In other words, some iterations observe less than $K$ distinct sets of violated constraints. This feature is analogous to observations in \cite{ng2018statistical,dekapowertech} on sparse set of active constraints in OPF.  
\subsection{Data-driven scenario enhancement}\label{sec:data-driven_enhance}
Note that while Section~\ref{sec:pf-aware_sel} allows us to select scenarios through prioritization metrics, we do not modify the generated scenarios. In this section, we present data-driven enhancements to selected scenarios before adding them to $\Omega_N$, that make our approach more efficient and amenable for large test-cases.
Based on preliminary tests on multiple cases, we observe that violations of a given constraint are primarily caused by a small subset of load fluctuations. Further, there are certain \emph{critical directions} for these load fluctuations that maximize violation. We now describe our method to identify these subset of loads and the critical directions, and a procedure to enhance the selected scenarios along these critical directions to make them more effective in enforcing feasibility. 
\subsubsection{Identifying critical components and directions}
Our approach to critical component identification relies on regularized linear regression \cite{wainwright2008graphical}, as described next. Consider selected scenario $t=(\mu^p(t),\mu^q(t))$ that we intend to enhance. Let $\mathcal{C}_{t}$ be the set of constraints violated by $t$ during recourse. For each $c\in\mathcal{C}_t$, let $\mathcal{S}_c$ be the set of random samples that violate it, with relative violation $u_c^s$ for sample $s = (\mu^p(s),\mu^q(s))$. We approximate a sparse linear map between the active and reactive loads fluctuations in buses $\mathcal{B}$, and violation for constraint $c\in \mathcal{C}_t$. The critical components and directions are identified via the vector $d_c$, computed as follows:
\begin{align*}
  &d_c = \arg\min_d\sum_{s\in\mathcal{S}_c}\left(u^s_c - (d_0 + \smashoperator[lr]{\sum_{\substack{i \in \mathcal{B}\\r=(p,q)}}} d^r_i\mu_i^r(s))\right)^2+\lambda\|d\|_1.
  \end{align*}
Here $\lambda>0$ is a regularization coefficient used with the $\ell_1$ norm to promote sparse solutions. This is an unconstrained convex optimization problem that can be easily solved, including in parallel for each selected scenario $t$ and constraint $c$.

\subsubsection{Scenario enhancement}
Using the critical directions identified, we describe the scenario enhancement procedure for the special case when the uncertainty is a uniform distribution over a box. There are variations possible for other distributions, which we do not pursue in the paper. The enhancement operation for scenario $t$ is given below:
\vspace{-2pt}
 \begin{align*}
    &\forall i \in \mathcal{B}, \ r=(p, q),\\
    & \textit{if } |d_i^r| < \tau_2 \textit{ then } \mu_i^r(t)\gets \mu_i^r(t)\\
    &\textit{else } \mu_i^r(t) \gets \left\{\begin{array}{ll}
    \overline{\mu_i^r} & \textit{if } d_i^r > \tau_2.\\\\
    \underline{\mu_i^r} & \textit{if } d_i^r < -\tau_2 
    \end{array}\right.
  \end{align*}
  where $\tau_2>0$ is a positive threshold.
  
Note that the enhancement step changes entries in scenario $t$ to their maximum or minimum values, based on the sign of non-trivial entries in $d_c$. This is done as the signs in $d_c$ reflect positive or negative directions to maximize violation. In settings where the maximum values of $\mu_i^p,\mu_i^q$ are not known, one can change it by a factor of the current entries (akin to a gradient based change). In this paper, we use $\tau_2 =1e-4$ for our simulations. By increasing the threshold $\tau_2$, the changes in $t$ can be made more sparse.

\subsubsection{Results for $73$-bus case} 
In addition to scenario selection of Section~\ref{sec:pf-aware_sel}, we now use the scenario enhancement technique on the $73$-bus test case. 
 The results are presented in Table \ref{tab:TE_extremization}. We observe that addition of scenario enhancement significantly reduces (more than $60\%$) the number of samples necessary for convergence of DDS-OPF. The combined impact of scenario selection and scenario enhancement steps over random sampling is evident from comparisons with Table \ref{tab:TE_results_first}. Using at most $11$ optimized scenarios, our proposed method is able to bring down infeasibility in out-of sample testing from $50\%$ to $0$. 
\begin{table}
  \centering
  \caption{Feasibility on $1,000$ out-of-sample scenarios for DDS-OPF with scenario selection \& scenario enhancement with $K=5$, for $73$-bus test system.}
  \begin{tabular}{|l|c|c|c|c|c|c|}
  \hline
  Policy & \# Iterations & $|\Omega_N|$ & $P^{1000}_{vio}$\\
  \hline
  MV & 1 & 6 & 0 \%\\
  \hline
  NC  & 2 & 11 & 0 \%\\
  \hline
  Hybrid  & 2 & 11 & 0 \%\\
  \hline
  \end{tabular}
  \label{tab:TE_extremization}
\end{table}

\subsection{Monte-Carlo step, Confidence bounds, and Scaling} \label{subsec:confidence_bounds}
In this section, we provide a theoretical confidence bound on the quality of the solution obtained from \algo{} based on the stopping criterion $\tau$ employed in step~$\mathbf{3}$. The proof relies on an application of the Hoeffding inequality \cite{wainwright2008graphical} and is omitted due to space considerations. 
\begin{theorem} \label{thm:large_deviation}
Suppose that for all nominal power flow solutions in $\Gamma_{OPF}$ and for all $\omega \in \Omega$, the  violation measure satisfies $|\mathcal{V}()| \leq M$. Then the solution $(p^0,v^0)$ obtained from \algo{} with stopping criterion $\tau$ and sample size $S$ satisfies
\begin{align*}   \label{eq:confidence_bound}
    \pw &\Bigg( \mathcal{SV} < \tau + \alpha S^{-1/2} \Bigg) \\
    &> 1- \delta,\quad\text{where}~~\alpha =  \sqrt{2M^2 \log(\small{1/\delta})}.
\end{align*}
\end{theorem}
{\begin{proof}
Since $\mathcal{V}(*,\omega)$ is a random variable as a function of uncertainty realization $\omega$ bounded by $M$ (the dependence on other non-random quantities has been suppressed for clarity). By using the Hoeffding inequality \cite{wainwright2008graphical} for \eqref{eq:stochastic_violation}, we get for any $t > 0$,
\begin{align*}
    \pw \left( \mathcal{SV} > \frac{1}{S}\sum_{i=1}^S \mathcal{V}(*,\omega_i) + t \right) \leq \exp\left(-St^2/2M^2 \right).
\end{align*}
The proof follows by using $t=\alpha S^{-1/2}$.
\end{proof}
}
Theorem~\ref{thm:large_deviation} shows how the stopping criterion translates to the quality of solution. A critical advantage of \algo{} is the Monte-Carlo-in-the-loop step $\mathbf{2}$. This is different from the vanilla scenario approach, where the random samples drawn from $\pw$ are incorporated into S-OPF. In contrast, in \algo{} the samples used in step $\mathbf{2}$ to evaluate the current solution $p^0,v^0$ are \emph{independent} from the samples used in the prior iteration to obtain $p^0,v^0$ (step~$\mathbf{5}$ or step~$1$ if first iteration). This results in fast convergence rates obtained via Theorem~\ref{thm:large_deviation}.

In all our experiments in Section~\ref{sec:numerical}, we choose $S = 1000$ and $\tau = 0$ with $\mathcal{SV} = $ probability of violation. Since the probability is always smaller than $1$, we have $M=1$. By applying Theorem~\ref{thm:large_deviation}, we can guarantee with confidence $95 \%$, that all solutions obtained in this paper satisfy the joint chance constraints with probability $99 \%$.

The Monte-Carlo step involves solving a series of power flows. Since the loading conditions resulting from uncertainty are still in the vicinity of the nominal load, warm-start methods can be used to solve a large number of power flows quickly. Further, this easily lends itself to parallelization, resulting in even further reduction in computation time. As a result, most of the computational complexity of \algo{} lies in solving the resulting S-OPF in step~$\mathbf{5}$.

\section{Numerical Experiments}
\label{sec:numerical}
In this section, we benchmark the \algo{} by detailed numerical experiments on a number of  test cases in the IEEE PES PGLib-OPF benchmark library. {The code is accessible from the following link : \url{https://github.com/imezghani/StochasticACOPF}.}

\subsection{Test cases and Experiment set up}
\begin{figure}
  \centering
  \includegraphics[scale=0.2]{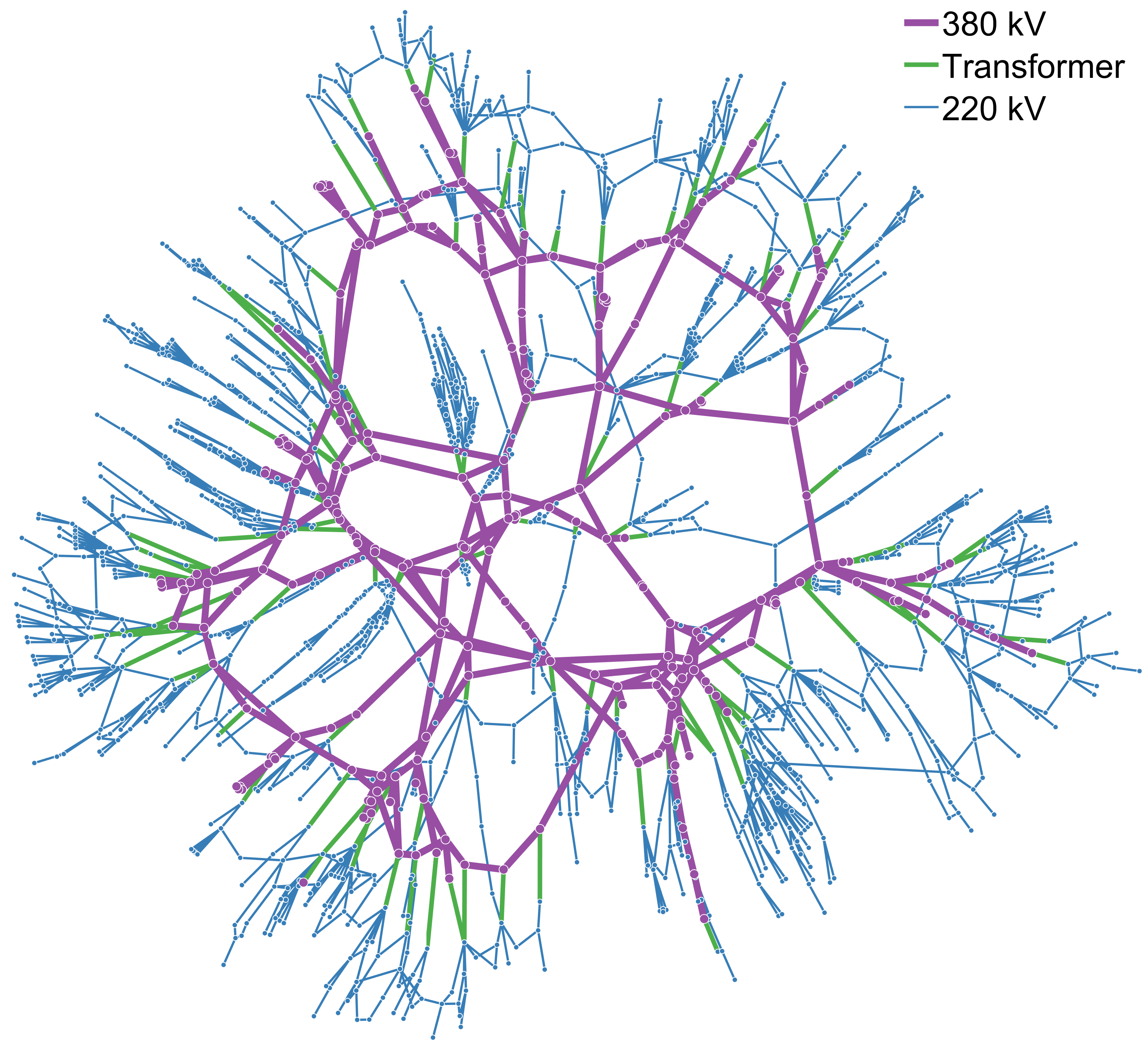}
  \caption{Topology of test case \texttt{1354\_pegase}.}
  \label{fig:topology1354}
\end{figure}

We consider four different test cases,  \texttt{24\_ieee}, \texttt{73\_ieee}, \texttt{118\_ieee} and \texttt{1354\_pegase}. The details of the test cases are shown in Table~\ref{tab:data}. For the first three (smaller) test cases, we assume that all active and reactive loads have a uniform $3\%$ fluctuation around their nominal value.
For the \texttt{1354\_pegase} test case, we assume that the real and reactive powers of the $211$ out of the $673$ loads that are situated at  end-buses fluctuate uniformly by $2\%$ of their nominal value. These buses often correspond to connections to distribution/sub-transmission, where the consumers and distributed energy resources responsible for the uncertainty are situated. The network is illustrated in Fig.~\ref{fig:topology1354}. From Table~\ref{tab:data}, it is clear that the recourse with the base-case solution can lead to infeasibility for an extremely high number of load fluctuations ($>85\%$). 

We remark here that the level of uncertainty chosen is quite large; increasing the uncertainty further from the given values makes a large percentage of loading conditions infeasible for the basic OPF, let alone the stochastic OPF.
For \algo{}, we choose $S = 1000$, $K = 5$ and $\tau = 0$ with empirical probability of violation  $\tilde{\mathcal{SV}}$.
\begin{table}[!htb]
  \centering
  \caption{Test case details}
  \label{tab:data}
 \resizebox{\columnwidth}{!}{\begin{tabular}{|l|c|c|c|c|c|c|}
  \hline
  Test case & \texttt{24\_ieee} & \texttt{73\_ieee} & \texttt{118\_ieee} & \texttt{1354\_pegase}\\
  \hline
  \# Buses & 24 & 73 & 118 & 1,354 \\
  \hline
  \# Generators & 33 & 99 & 54 & 260\\
  \hline
  \# Lines & 38 & 120 & 186 & 1,991\\
  \hline
  \# Loads & 17 & 51 & 99 & 673\\
  \hline
  \# Fluctuations & 17 & 51 & 99 & 211\\
  \hline
  Base cost & $6.34e4$ & $1.90e5$ & $9.72e4$ & $1.26e6$\\
  \hline
  Base $P^{1000}_{vio}$ & 87.5 \% & 100\% & 100\% & 100\%\\
  \hline
  \end{tabular}
  }
\end{table}

\subsection{Performance trends} \label{subsec:performance_trends}
Table~\ref{tab:results_small} shows the results of applying \algo{} on the different test cases.
\subsubsection{Scenario size}
We see that \algo{} has excellent performance for all test cases in terms of number of iterations ($\# It$) and final number of samples ($|\Omega_N|$). $|\Omega_N|$ grows very slowly with network size, from $7$ on the $24$ bus system to only $31$ on the $1354$ bus system. This demonstrates that \algo{} has very favorable scaling properties, and can be scaled to even larger systems. 
\subsubsection{Cost} While the scenario enhancement procedure introduced in Section~\ref{sec:data-driven_enhance} helps quickly obtain a secure solution, worsening the scenarios can potentially increase the cost. However, we see by comparing the cost between Table~\ref{tab:data} and Table~\ref{tab:results_small} that in all the cases the increase in cost from the deterministic (and unsafe) solution is small with $\sim2\%$ for the $24$ bus system to $\sim 0.2\%$ for the $1354$ bus system. The larger, more realistic system, possesses more flexibility to handle uncertainty in a more economic way, as expected.\\
{Furthermore, by comparing the costs for the $73$-bus system in Tables \ref{tab:TE_results_first} and  \ref{tab:results_small}, it is clear that our algorithm achieves the same cost as the vanilla scenario selection scheme, while significantly improving the feasibility of the solution.}

{\subsubsection{Distance to the deterministic solution} The last two columns of Table \ref{tab:results_small} report the $2-$norm difference between the deterministic solution set-points and the \algo{} solution set-points, first in terms of real power injections and then voltage magnitudes. The quantities suggest that the solution to the stochastic OPF lies in the vicinity of the solution of the deterministic OPF. Nevertheless, this adjustment to the deterministic solution is critical and can significantly improve the robustness of the solutions. Using the $1354$ bus system as an example, the reduction in maximum violation can be as much as $~17\%$ (see Table~\ref{tab:results_1354}).}


\begin{table}
  \centering
  \caption{Overall performance trends of \algo{}}
  \label{tab:results_small}
 \resizebox{\columnwidth}{!}{ \begin{tabular}{|l|c|c|c|c|c|c|c|c|}
  \hline
  Test case &  Policy & \# It & $|\Omega_N|$ & $P^{1000}_{vio}$ & Cost & {Dist P} & {Dist V}\\
  \hline
\texttt{24\_ieee} &{MV}  & 3 & 7 & 0 \%  & 6.502e4 & {$2.5e{-1}$} & {$4.4e{-3}$}\\
  \hline
\texttt{73\_ieee} & {MV}  & 1 & 6 & 0 \% & 1.948e5 &{$5.4e{-1}$} & {$9.9e{-3}$}\\
  \hline
\texttt{118\_ieee} &  {Hybrid}  & 3 & 14 & 0 \% & 9.802e4 &{$9.0e{-1}$} & {$3.7e{-2}$}\\
  \hline
\texttt{1354\_pegase} & {MV} & 6 & 31 & 0.1 \% &  1.263e6 & {$1.4$} & {$5.1e{-2}$}\\  
\hline
  \end{tabular}}
\end{table}


\subsection{A detailed study on the \texttt{1354\_pegase} test case}
We present detailed numerical experiments for different variants of \algo{} on the $1354$ bus system.
Table \ref{tab:results_1354} shows the results  for different choices of prioritization rule. 
\subsubsection{Effect of prioritization rule}
With any prioritization rule, \algo{} finds an excellent solution with a maximum of $31$ scenarios in the final S-OPF. All resulting costs are similar, and within $0.2 \%$ of the base case cost.
\subsubsection{Different stochastic violation measures}
The Monte Carlo in-the-loop method employed by \algo{} grants it the flexibility to handle a variety of stochastic violation measures. Table~\ref{tab:results_1354} shows two such violation measure, the \emph{probability of violation} and \emph{maximum magnitude of violation} in an out of sample testing with $1000$ samples. This translates into confidence guarantees in the sense of Theorem~\ref{thm:large_deviation}.
As an example by using Theorem~\ref{thm:large_deviation}, we can guarantee that the solution obtained using $MV$ for constraint selection, satisfies a chance constraint with probability of violation $<1.1 \%$. Similarly, we can guarantee that  in the face of uncertainty, the solution has a maximum constraint violation of $3.26 \%$. The second guarantee uses a very conservative maximum violation bound of $M=10$. Both the above statements carry a confidence of $95\%$.

\begin{table}
  \centering
    \caption{Results of the iterative approach on \texttt{1354\_pegase}.}
  \label{tab:results_1354}
 \begin{tabular}{|l|c|c|c|c|c|c|}
  \hline
  Policy &  \# It & $|\Omega_N|$ & $P^{1000}_{vio}$ & Max. Viol. &  Cost ($\times1e6$)\\
  \hline
  Base & - & 1 & 100 \% & 17.3 \% &  1.2620\\
  \hline
  MV &  6 & 31 & 0.1 \% & 0.06 \% &  1.2633\\  
  \hline
  NC &  6 & 31 & 2.3 \% & 0.34 \% &  1.2633 \\  
  \hline
  Hybrid &  8 & 31 & 0.1 \% & 0.04 \% & 1.2634\\  
  \hline
\end{tabular}
\end{table}

\section{Conclusion and Future directions}
This paper describes a principled iterative data-driven approach for stochastic AC-OPF under general probabilistic constraints. The non-linear and non-convex equations in AC-OPF make random sampling or scenario reduction approaches impractical for large test cases, due to their large sample requirement. Our data-driven algorithm is able to overcome that by a novel 2-step process for `dominant' scenario design/construction that involves: (a) scenario selection based on constraint violations, and (b) scenario enhancement by regularized linear regression. Through system-level intuition, theoretical bounds, and finally numerical verification on multiple test cases, we demonstrate that our data-driven algorithm is able to provide feasible solutions to stochastic AC-OPF using far lower scenarios than conventional schemes. For example, our method uses only $31$ constructed samples to provide a feasible solution for the \texttt{1354\_pegase} test case, that  satisfies chance constraints with $<1.1\%$ violation probability.

This work naturally leads to multiple extensions. First, we want to parallelize the steps (scenario enhancement, Monte Carlo checks) and include warm-starts in our algorithm to achieve its true computational benchmark. While the current work operates on box-uncertainty sets for sampling and scenario enhancement, efficient data-driven efforts for general (non-parametric) uncertainty sets is another direction for exploration. Finally we plan to analyze extensions of our approach to related and computationally challenging problems on resilient network design and stochastic unit commitment.



\bibliographystyle{IEEEtran}
\bibliography{biblio.bib}
%



\end{document}